\documentstyle[12pt,amssymb,amstex]{amsart}
\newtheorem{tm}{\\ $\ \ \ \ $ Theorem}
\newtheorem{lm}{\\ $\ \ \ \ $ Lemma}

\newcounter{rem}

\begin{document}
\title{On the Lieb-Thirring constants $L_{\gamma,1}$ for $\gamma\geq 1/2.$}
\author{Timo Weidl}
\date{April 2, 1995}
\address{Timo Weidl\\Royal Institute of Technology\\Department of Mathematics\\
S-10044 Stockholm\\Sweden\\int+46-8-7906196, fax: int+46-8-7231788}
\email{weidl@@math.kth.se}
\maketitle
\begin{abstract}
Let $E_i(H)$ denote the negative eigenvalues of the one-dimensional
Schr\"odinger operator $Hu:=-u^{\prime\prime}-Vu,\ V\geq 0,$
on $L_2({\Bbb R}).$ We prove the inequality
\begin{equation}\label{Wei}
\sum_i|E_i(H)|^\gamma\leq L_{\gamma,1}\int_{\Bbb R} V^{\gamma+1/2}(x)dx,
\end{equation}
for the "limit" case $\gamma=1/2.$ This will imply improved estimates
for the best constants $L_{\gamma,1}$ in \eqref{Wei}
as $1/2<\gamma<3/2.$\end{abstract}

{\bf 0.} Let $H=-\Delta-V$ denote the Schr\"odinger operator
 in $L_2({\Bbb R}^d).$
If the potential $V\geq 0$ decreases sufficiently fast at infinity,
the negative part of the spectrum of
$H$ is discrete. Let $\{E_i(H)\}$ be the corresponding
increasing sequence of negative eigenvalues, each eigenvalue occurs
with its multiplicity. This sequence is either finite or tends to zero.

Estimates on the behavior of the sequence of eigenvalues in terms of the
potential have been in the focus of research for many years. In the earlier
papers the main attention was paid to bounds on the number of negative
eigenvalues ( \cite{Barg},\cite{B},\cite{Schwinger},\cite{Rbl},
\cite{C},\cite{[L]},
\cite{LiYau},\cite{Blanch} ).
In \cite{LT} Lieb and Thirring proved
inequalities of the type
\begin{equation}\label{LiTh}
\sum_i|E_i(H)|^\gamma\leq L_{\gamma,d}\int_{{\Bbb R}^d}
V^{\gamma+\varkappa}(x)dx,
\ \ \varkappa=d/2.
\end{equation}
Since that these estimates
and the corresponding constants $L_{\gamma,d}$ have
been studied intensively (e.g. \cite{Lieb},\cite{GGM},\cite{RoHe}).
Up to now it was known, that \eqref{LiTh} holds for all $\gamma\geq 0$
if $d\geq 3,$ for $\gamma>0$ if $d=2,$ and for $\gamma>1/2$ if $d=1.$
On the contrary \eqref{LiTh} fails for $\gamma=0,
d=2$ and for $\gamma<1/2, d=1.$
In this paper we prove \eqref{LiTh} for the remaining case $d=1,\gamma=1/2,$
which does not seem to have been settled so far.
This result will imply an essential improvement for the estimates
on the constants $L_{\gamma,1},\ 1/2<\gamma<3/2.$
Moreover we deduce a new integral bound on the transmission coefficient
of the corresponding scattering problem.

In conclusion the author expresses his gratefulness to M. Sh. Birman,
who introduced him to the topic of negative bound states of Schr\"odinger
operators. Moreover I am grateful to A. Laptev, under whose intensive
supervision this paper was written.
\vspace{3mm}

{\bf 1.} In this subsection we provide
some auxiliary results on the negative spectrum
of the Neumann problem for the Sturm-Liouville-operator
\[(L_I^Nu)(x)=-u^{\prime\prime}(x)-V(x)u(x),\]
\[x\in I=[0,l],
\ \ u^\prime(0)=u^\prime(l)=0,\ \ 0\leq V(x)\in L_1(I).\]
Let $N_I(V,E)$ be the number of eigenvalues
$E_i(L_I^N)$ of $L_I^N$ below $E<0.$
According to
the Birman-Schwinger principle (\cite{B},\cite{Schwinger}),
the value of $N_I(V,E)$ does not exceed
the square of the Hilbert-Schmidt norm of the integral operator
\[(Q_E u)(x)=\sqrt{V(x)}\int_0^l G(x,y,E)\sqrt{V(y)}u(y),\ \ x\in I.\]
Here
\[G(x,y,E)=\Biggl\{ \begin{array}{cc}
\frac{\cosh(\lambda x)\cosh(\lambda(y-l))}{\lambda\sinh(\lambda l)}\ \ x\leq y
\\
\frac{\cosh(\lambda y)\cosh(\lambda(x-l))}{\lambda\sinh(\lambda l)}\ \ y\leq x
\end{array}\Biggr . ,\ \ \lambda=\sqrt{|E|},\ E<0,\ x,y\in I,\]
denotes the Green function of the problem
$-u^{\prime\prime}-Eu,\ u^\prime(0)=u^\prime(l)=0$ on $I.$
In view of
\[|G(x,y,E)|\leq \frac{\coth(\lambda l)}{\lambda}\]
one obtains the inequality
\begin{equation}\label{N}
N_I(V,E)\leq \frac{\coth^2(\lambda l)}{\lambda^2}
\biggl(\int_I V(x)dx\biggr)^2,\ \ \lambda=\sqrt{|E|},\ E<0.
\end{equation}
We apply \eqref{N} to the lowest eigenvalue $E_1(L_I^N),$
and find
\begin{equation}\label{N1}
\vartheta (\lambda_1 l)  \leq l \int_I V(x)dx,
\ \ \lambda_1=\sqrt{ |E_1(L_I^N)| }>0,
\ \ \vartheta(x):=x\tanh x.
\end{equation}
The function $\vartheta(x)=x\tanh x$
is strongly increasing in $x\geq 0.$ Let
$\varsigma(y)$ be the inverse function of
$\vartheta(x)=y,\ x,y>0.$ From \eqref{N1} we immediately conclude

\begin{lm}\label{lm1}
 Let $E_1(L_I^N)$ be the lowest eigenvalue of the
Neumann problem $L_I^N$ on $I=[0,l].$ Assume $0\leq V\in L_1(I).$
Then the estimate
\begin{equation}\label{N1lm}
\lambda_1\leq \varsigma(l \int_I V(x)dx)/l,\ \ \
\ \ \lambda_1=\sqrt{ |E_1(L_I^N)| }\geq 0,
\end{equation}
holds.
\end{lm}
\vspace{3mm}

Next we recall a criteria, providing the existence of not more then
one negative eigenvalue of the operator $L_I^N.$

First notice, that for
functions $u\in C^\infty(I),$ satisfying the
orthogonality condition $\int_I udx=0,$ the inequality
\begin{equation}\label{3}
|u(x)|^2\leq \frac{l}{3}\int_I|u^\prime|^2dx,\ \ x\in I,
\end{equation}
holds. Indeed, we have
\[lu(x_0)=\int_0^{x_0}u^\prime (x)xdx-\int_{x_0}^lu^\prime (x)(l-x)dx.\]
This gives
\[|u(x_0)|^2\leq \frac{(x_0^{3/2}+(l-x_0)^{3/2})^2}{3l^2}
\int_0^l |u^\prime (x)|^2dx .\]
Passing to the upper bound in $x_0\in I$ we find \eqref{3}.
The constant $l/3$ in \eqref{3} is sharp.

\begin{lm}\label{lm2}
Assume, that for the non-trivial potential $0\leq V$ the estimate
\begin{equation}\label{NN}
l\int_I V(x)dx \leq 3
\end{equation}
holds. Then the Neumann problem $L_I^N$ on $I=[0,l]$ has exactly
one negative eigenvalue.
\end{lm}

{\bf Proof.} The existence of the eigenvalue is obvious.
By \eqref{3} we find
\begin{equation}\label{ininin}
\int_I |u^\prime|^2dx-\int_I V(x)|u|^2dx \geq 0,
\ \ u\in C^\infty([0,l]),\ \int_I udx=0.
\end{equation}
The inequality \eqref{ininin} holds on a set of functions of codimension
one with respect to the domain of the quadratic form of the
Neumann problem $L_I^N.$ Thus $L_I^N$ itself has not more than one
negative eigenvalue. $\Box$
\vspace{3mm}

{\bf 2.} We turn now our attention to the one-dimensional Schr\"odinger
operator
\[Hu=-u^{\prime\prime}-V(x)u,\ \ x\in{\Bbb R},
\ 0\leq V\in L_1({\Bbb R}),\]
realized as a self-adjoint operator on $L_2({\Bbb R})$ in the form sum sense.
Let $H_+$ and $H_-$ denote selfadjoint operators on $L_2({\Bbb R}_\pm),$
corresponding to the Neumann problem on
the positive and negative semi-axes respectively.

Assume $V\not\equiv 0$ on ${\Bbb R}_+.$
Fix the point $l_0=0,$ and by iteration construct the sequence $l_k,
k\in{\Bbb K}\subset{\Bbb N},$
\begin{equation}\label{NNcon}
 l^{(k)}\int_{l_k}^{l_{k+1}}V(x)dx=3,\ \  l^{(k)}:=l_{k+1}-l_k.
\end{equation}
If it occurs that $\int_{l_n}^\infty V(x)dx=0,$
we formally choose $l_{n+1}=+\infty.$
For the elements of the sequence $l^{(k)}$ we have
the bound $l^{(k)}\geq 3/\int V(x)dx>0.$ Hence
the intervals $I_k:=[l_k,l_{k+1}],\ k\geq 0,$
cover ${\Bbb R}_+.$

On each interval we consider the Neumann problem
$L_{I_k}^Nu=-u^{\prime\prime}-V(x)u,\ u^\prime(l_k)=u^\prime(l_{k+1})=0.$
Let $H_+^N=\oplus_{k\in{\Bbb K}} L_{I_k}^N$ denote the orthogonal sum
of these operators.
We have  $H_+^N\leq H_+.$ For the ordered sequence of the respective
negative eigenvalues this implies
\begin{equation}\label{specpro}
E_i(H_+^N)\leq E_i(H_+).
\end{equation}
In case of a semi-infinite interval the potential is identically
zero on this interval, the respective Neumann problem
has no negative spectrum. Therefore it will not play any role in our
considerations.

By Lemma \ref{lm2} the Neumann problem
$L_{I_k}^N$ on the finite intervals
$I_k$ has exactly one negative eigenvalue.
Because of \eqref{NNcon} the bound \eqref{N1lm}
for
$\lambda_{1}(I_k):=\sqrt{|E_{1}(L_{I_k}^N)|}$ turns into
$\lambda_1(I_k)\leq \varsigma(3)/l^{(k)},$ or equivalently
\footnote{On the other hand for $u(x)=1/\sqrt{l}$ one has
$E_1(L_{I_k}^N)\leq (L_{I_k}^Nu,u)_{L_2(I_k)}=-l^{-1}\int_{I_k}V(x)dx,$
and $\lambda_1(I_k)\geq \sqrt{1/3}
\int_{I_k} V(x)dx.$}
\begin{equation}\label{victoria}
\lambda_1(I_k)\leq \frac{\varsigma(3)}{3}
\int_{I_k} V(x)dx.\end{equation}
Since $V\in L_1({\Bbb R}_+),$
the sequence $\int_{I_k} V ( x )dx$ tends to zero as $k\to\infty.$
Thus both operators $H_+^N$ and $H_+$ are semibounded
and their negative spectra are discrete.
The negative spectrum of $H_+^N$ coincides (as set
and in its multiplicity) with the sequence of eigenvalues
$\{E_1(L_{I_k}^N)\}=\{-\lambda_1^2(I_k)\}.$
By \eqref{specpro} we have
$|E_i(H_+)|\leq |E_i(H_+^N)|.$ Together with
$0\leq V\in L_1({\Bbb R}_+)$ this implies
\[\sum_i \sqrt{|E_i(H_+)|}\leq\sum_i \sqrt{|E_i(H_+^N)|}=\sum_{k}
\lambda_{1}(I_k)\leq\]
\[\leq \frac{\varsigma(3)}{3}\sum_k \int_{I_k} V(x)dx=
\frac{\varsigma(3)}{3}\int_0^\infty V(x)dx,\]
and we find the claimed result
for the negative eigenvalues of the Neumann operator
on the semi-axes
\begin{equation}\label{upp+-}
\sum_i \sqrt{|E_i(H_+)|}\leq L_{\frac{1}{2},1}^+
\int_{{\Bbb R}_+} V(x)dx,
\ \ L^+_{\frac{1}{2},1}\leq \varsigma(3)/3<1.005.
\end{equation}
Naturally the analogous bound with the same constant
holds for the operator $H_-.$
Because of $H_-\oplus H_+\leq H$ we obtain
the analog estimate on the negative eigenvalues
of the Schr\"odinger operator $H$ on ${\Bbb R}$
\begin{equation}\label{upp}
\sum_i \sqrt{|E_i(H)|}\leq L_{\frac{1}{2},1}
\int_{{\Bbb R}} V(x)dx,
\ \ L_{\frac{1}{2},1}\leq \varsigma(3)/3<1.005.
\end{equation}

We recall the reverse estimate
for the operator $H$ (see \cite{LT}
and \cite{GGM}).
The first sum rule of Faddeev-Zakharov \cite{FadZ} states
\begin{equation}\label{FadZ}
\int V(x)dx = 4\sum_i \sqrt{|E_i(H)|}+\pi^{-1}\int \ln(1-|R(k)|^2)dk,
\end{equation}
for (not necessary sign-defined)
potentials $V\in C_0^\infty({\Bbb R}).$
In this $R(k)\in [0,1]$ is the reflection coefficient of the operator $H.$
The integrand on the right hand side is negative, hence
\begin{equation}\label{FadZ1}
\sum_i \sqrt{|E_i(H)|}\geq\frac{1}{4}\int V(x)dx.
\end{equation}
This bound can be closed to all potentials $V\in L_1({\Bbb R})$.

The estimate from below on the constant $L_{1/2,1}$ can be improved.
For a potential
$0\leq V\in L_1({\Bbb R})$ the number $N(V,E)$ of eigenvalues $E_i(H)<E<0$
is bounded by
\[N(V,E)\leq \frac{1}{2\sqrt{|E|}}\int Vdx,\]
(see (3.7) in \cite{BS2}). For the lowest eigenvalue this gives
\begin{equation}\label{BSW}
\sqrt{|E_1(H)|}\leq \frac{1}{2} \int Vdx.
\end{equation}
The constant in this estimate is sharp. Indeed,
if the non-trivial potential $0\leq V\in C_0^\infty({\Bbb R})$ is
supplied with a sufficiently small coupling constant $\alpha>0,$
the operator $H_\alpha u=-u^{\prime\prime}-\alpha Vu$ has exactly one
negative  eigenvalue $E_1(H_\alpha).$
This eigenvalue obeys the asymptotics (see \cite{Si})
\[\sqrt{|E_1(H_\alpha)|}= ( \alpha/2 )\int Vdx+o ( \alpha ),\ \ \alpha\to 0.\]
We conclude $L_{1/2,1}\geq 1/2.$

The previous arguments can be adapted to the problem on the semi-axes.
Assume, that $0\leq V$ is continuous on ${\Bbb R}_+$ up to the point
zero, and has compact support. We supply this potential with a small
coupling constant $\alpha>0,$ and consider the lowest
eigenvalue $E_1(H_{+,\alpha})$ of the respective Neumann problem on
${\Bbb R}_+.$ Let $u_\alpha(x)$ denote the corresponding eigenfunction.
The even extension $u_\alpha(x)=u_\alpha(-x)$ is an eigenfunction
of the operator $H_\alpha$ with the extended potential $V(x)=V(-x)$ on
${\Bbb R}.$ The corresponding eigenvalue is
$E_1(H_\alpha)=E_1(H_{+,\alpha}).$ Since the operators $H_\alpha$ and
$H_{+,\alpha}$ have only one negative eigenvalue for sufficiently small
$\alpha>0,$ we find
\[\sqrt{|E_1(H_+)|}=\alpha\int_0^\infty Vdx+o ( \alpha ),\ \ \alpha\to 0.\]
We obtain $1\leq L_{1/2,1}^+<1.005,$ our bound on the constant
for the Neumann problem on the semi-axes is almost sharp!

Finally we remark the analog of \eqref{FadZ1}
for the operator $H_+.$ For a sumable potential
$V(x)=V(-x)$ it holds
\begin{equation}\label{FadZ2}
\int_{0}^\infty V(x)dx\leq
2\sum_i\sqrt{|E_i(H)|}\leq
2 \sum_i\sqrt{|E_i(H_-\oplus H_+)|}=4\sum_i\sqrt{|E_i(H_+)|}.
\end{equation}

The results of this subsection we summarize in

\begin{tm}\label{thethe}

1.) The inclusion $0\leq V\in L_1({\Bbb R}_+)$ implies
the inequality
\begin{equation}\label{upp0}
\sum_i \sqrt{|E_i(H_+)|}\leq L^+_{\frac{1}{2},1}\int_0^\infty V(x)dx.
\end{equation}
For the best constant $L^+_{1/2,1}$ in \eqref{upp0}
we have the estimate $1\leq L_{1/2,1}^+\leq \varsigma(3)/3<1.005.$
Reversely, a priori assuming $0\leq V\in L_1^{loc}({\Bbb R}_+),$
the discreteness of the negative spectrum together
with the convergence of the sum in \eqref{upp0} imply $V\in L_1$ and
\eqref{FadZ2}.

2.) The inclusion $0\leq V\in L_1({\Bbb R})$ implies
the inequality
\begin{equation}\label{upp1}
\sum_i \sqrt{|E_i(H)|}\leq L_{\frac{1}{2},1}\int V(x)dx.
\end{equation}
For the best constant $L_{1/2,1}$ in \eqref{upp1}
we have the estimate $1/2\leq L_{1/2,1}\leq \varsigma(3)/3<1.005.$
Reversely, a priori assuming $0\leq V\in L_1^{loc}({\Bbb R}),$
the discreteness of the negative spectrum together
with the convergence of the sum in \eqref{upp1} imply $V\in L_1$ and
\eqref{FadZ1}. \end{tm}

{\bf Remark.} As usually one can drop the assumption $V\geq 0.$ One has
to ensure, that the corresponding operators $H,H_+$ are defined in the
form sum sense, and the integrant in \eqref{upp0} and \eqref{upp1} has
to be replaced by $V_+(x):=\max\{0,V(x)\}.$

Notice, that \eqref{upp1} and \eqref{FadZ} together with
$1/2\leq L_{1/2,1}<\infty$ imply

\begin{tm} Assume $V\in C_0^\infty({\Bbb R}),
2V_\pm=|V|\pm V,$ and let $R(k)$
be the reflection coefficient for the corresponding one-dimensional
Schr\"odinger operator $Hu=-u^{\prime\prime}-Vu$ on $L_2({\Bbb R}).$
Then the integral estimate
\[\frac{1}{\pi}
\int |\ln (1-|R(k)|^2)| dk \leq \int V_-dx+(4L_{1/2,1}-1)\int V_+dx
\leq(4L_{1/2,1}-1)\|V\|_{L_1({\Bbb R})}\]
holds.
\end{tm}

\vspace{3mm}

{\bf 3.} We turn now to the case $\gamma>1/2.$ We restrict our
considerations to the operator $H$ on $L_2({\Bbb R}).$
Here the inequalities
\begin{equation}\label{Wei12}
\sum_i|E_i(H)|^\gamma\leq L_{\gamma,1}\int_{\Bbb R} V^{\gamma+1/2}(x)dx,
\end{equation}
are well established,
but we will give an essential improvement of the
estimates for the corresponding constants $L_{\gamma,1}.$ For
$\gamma\geq 3/2$ in \cite{AiL} has been proven,
that $L_{\gamma,1}=L_{\gamma,1}^{cl}.$
The last notation stands for the classical constant
\[ L_{\gamma,1}^{cl}=\frac{\Gamma(\gamma+1)}{2\sqrt{\pi}
\Gamma(\gamma+\frac{3}{2})}.\]
Hence we will stress on the case $1/2<\gamma<3/2.$ We shall compare
our results with the bounds of Lieb and Thirring
\begin{equation}\label{LiThCo}
L_{\gamma,1}\leq L_{\gamma,1}^{LT} :=\frac{\gamma^{\gamma+1}}
{\sqrt{2}(\gamma-1/2)^{\gamma+1/2}(\gamma+1/2)},
\end{equation}
and their improvements by Glaser, Grosse and Martin (\cite{GGM})
\begin{equation}\label{GGM1}
L_{\gamma,1}\leq L_{\gamma,1}^{GGM}:=\inf_{1<m<3/2}
\frac{(m-1)^{m-1}\Gamma(2m)\gamma^{\gamma+1}\Gamma(\gamma+\frac{1}{2}-m)}
{2^{2m-1}m^{m-1}\Gamma(m)\Gamma(\gamma+\frac{3}{2})
(m-\frac{1}{2})^{m-\frac{1}{2}}(\gamma+\frac{1}{2}-m)^{\gamma+\frac{1}{2}-m}}.
\end{equation}

Our proof of Theorem \ref{thethe}
can be generalized to the case $\gamma\geq 1/2.$
However, this direct approach gives the bound
$L_{\gamma,1}\leq(\varsigma(3))^{2\gamma}/3^{\gamma+1/2},$
which is not very sharp.
A better bound can be found
using the fact, that the ratio $L_{\gamma,1}/L_{\gamma,1}^{cl}$ is
non-increasing in $\gamma,$
see \cite{AiL}.
We find
\[L_{\gamma,1}\leq L_{\gamma,1}^*:=4\varsigma(3)L_{\gamma,1}^{cl}/3=
\frac{2\varsigma(3)\Gamma(\gamma+1)}{3\sqrt{\pi}
\Gamma(\gamma+\frac{3}{2})}.\]
This bound is sharper than \eqref{LiThCo} and \eqref{GGM1} for all
$1/2\leq\gamma\leq 3/2.$
In particular, $L_{1,1}^{*}<0.853\ ,$
while $L^{LT}_{1,1}=4/3$ and $L_{1,1}^{GGM}=1.269.$
\footnote{One can apply an argument of Glaser, Grosse and
Martin \cite{GGM}, to deduce
a bound on $L_{0,3}^{sph}$ for {\em spherical symmetric} potentials
from $L_{1,1}$ .
Although one considers only a special class of potentials,
even the new bound on $L_{1,1}$ is not sharp enough to reach
Lieb's result for $L_{0,3}$ by this method.}

If we consider only potentials $\tilde{V}$
proportional to a characteristic function
of a set $M\subset{\Bbb R}$ of finite measure
\[\tilde{V}(x)=v\chi_M(x),\ \ \chi_M(x)=\Bigl\{
\begin{array}{cc}1\ \ x\in M\\0\ \ x\not\in M\end{array},
\ \ v>0,\]
we can find a better constant by
"interpolating" between
the cases $\gamma=1/2$ and $\gamma=3/2.$
Indeed, the ratio
\[\psi(\gamma,\tilde{V}):=\frac{\sum_i |E_i(H)|^\gamma}
{\int \tilde{V}^{\gamma+1/2}dx}\]
is analytic and continuous up to the boundary for complex $\gamma$ in the
strip $1/2<\Re\gamma<3/2.$ On the boundary we have the estimates
\[|\psi(\gamma,\tilde{V})|
\leq L_{1/2,1}\leq \varsigma(3)/3,\ \ \mbox{as}\ \Re\gamma=1/2,
\ \ |\psi(\gamma,\tilde{V})|
\leq L_{3/2,1}=3/16,\ \ \mbox{as}\ \Re\gamma=3/2.\]
By the Hadamard Lemma we obtain
\begin{equation}\label{sharr}\psi(\gamma,\tilde{V})\leq
\tilde{L}_{\gamma,1}^{**}
:=\bigl(\frac{\varsigma(3)}{3}\bigr)^{\frac{3}{2}-\gamma}
\bigl(\frac{3}{16}\bigr)^{\gamma-\frac{1}{2}},\ \ 1/2<\gamma<3/2.
\end{equation}
In particular, $\tilde{L}_{1,1}^{**}<0.4341.$ We notice, that
\eqref{sharr} is sharper than the results
for characteristic functions by A. Laptev in \cite{La}
for the
case of dimension one .

For completeness we recall the estimate from below
on the constants $L_{\gamma,1},$
obtained in \cite{LT}.
To do so we consider the best
constants $L_{\gamma,1}^1$ in the inequalities
\begin{equation}\label{1//2}
|E_1(H)|^\gamma\leq L_{\gamma,1}^1 \int V^{\gamma+1/2}dx,\ \ \gamma\geq 1/2.
\end{equation}
Obviously $L_{\gamma,1}\geq L_{\gamma,1}^1.$
For $\gamma>1/2$ the corresponding variational equation can be solved
analytically and one obtains
\footnote{In particular this gives $0.2451<L_{1,1}<0.853.$}

\begin{equation}\label{L!}
L_{\gamma,1}^1= \pi^{-1/2}\frac{1}{\gamma-1/2}
\frac{\Gamma(\gamma+1)}{\Gamma(\gamma+1/2)}
\biggl(\frac{\gamma-1/2}{\gamma+1/2}\biggr)^{\gamma+1/2}
=2L_{\gamma,1}^{cl}\biggl(\frac{\gamma-1/2}{\gamma+1/2}\biggr)^{\gamma-1/2},
\end{equation}
(see \cite{LT}). Moreover in the previous subsection we showed, that \eqref{L!}
remains true for $\gamma=1/2$ and $L_{1/2,1}^1=1/2.$
For $\gamma\geq 3/2$ it holds $L_{\gamma,1}^1\leq L_{\gamma,1}^{cl}.$ For
$\gamma<3/2$ we have $L_{\gamma,1}^1>L_{\gamma,1}^{cl},$ this implies
$L_{\gamma,1}>L_{\gamma,1}^{cl}$ as $1/2\leq \gamma<3/2$ (see \cite{LT}
and also \cite{RoHe}).

We proved

\begin{tm} For the numerical values of the
best possible constants $L_{\gamma,1},\ 1/2\leq\gamma\leq 3/2$
in \eqref{Wei12} the estimate
\[2L_{\gamma,1}^{cl}\biggl(\frac{\gamma-1/2}{\gamma+1/2}\biggr)^{\gamma-1/2}
\leq L_{\gamma,1}\leq L_{\gamma,1}^{*}=
4\varsigma(3)L_{\gamma,1}^{cl}/3=
\frac{2\varsigma(3)\Gamma(\gamma+1)}{3\sqrt{\pi}
\Gamma(\gamma+\frac{3}{2})},
\ \ \frac{1}{2}\leq\gamma\leq\frac{3}{2},\]
holds. For potentials $\tilde{V}$
proportional to characteristic functions, the
constant $L_{\gamma,1}$
in the Lieb-Thirring inequality
can be replaced by $\tilde{L}_{\gamma,1}^{**}$ from \eqref{sharr}.
\end{tm}

Notice, that the bound $L_{\gamma,1}^*$ on $L_{\gamma,1}$ does not
tend to $L_{\frac{3}{2},1}=L_{\frac{3}{2},1}^{cl}=3/16$ as $\gamma\to
3/2-0.$
For $\gamma$ near $3/2$
the estimate on $L_{\gamma,1}$ can be improved. To do so we
shall recall some auxiliary material from real interpolation theory.
\vspace{3mm}

{\bf 4.} Let $\ell_p$ denote the ideal of $p$-sumable sequences
$\{u_n\}_{n\in{\Bbb N}},$ equipped by the standard quasi-norm
\[\|\{u_n\}\|_{\ell_p}^p:=\sum_n |u_n|^p\ ,\ \ p>0.\]
For
a sequence $\{u_n\}_{n\in{\Bbb N}}\in \ell_{p_0}+\ell_{p_1}$
one can define the $(p_0,p_1)-K$-function
\[K(\{u_n\},t,p_0,p_1):=\inf_{
\begin{array}{cc}u_n=u_n^{(0)}+ u_n^{(1)}\\u_n^{(i)}\in\ell_{p_i}
\end{array}}\Bigl(\|u_n^{(0)}\|_{\ell_{p_0}}^{p_0}
+t\|u_n^{(1)}\|_{\ell_{p_1}}^{p_1}
\Bigr),\ \ t>0.\]
For a function $f\in L_{p_0}+L_{p_1}$ one may use the analogous
definition
\[K(f,t,p_0,p_1):=\inf_{
\begin{array}{cc}f=f_0+ f_1\\f_i\in L_{p_i}
\end{array}}\Bigl(\|f_0\|_{L_{p_0}}^{p_0}
+t\|f_1\|_{L_{p_1}}^{p_1}
\Bigr),\ \ t>0.\]

On functions $h:(0,\infty)\to [0,\infty)$ we define the functionals
\[\Phi_{\eta,q}[h]=
\Bigl(\int_0^\infty(t^{-\eta}h(t))^q\frac{dt}{t}\Bigr)^{1/q},
\ \ \eta\in(0,1),\ 0<q<\infty,\]
\[\Phi_{\eta,\infty}[h]=\sup_{t>0}t^{-\eta}h(t),
\ \ \eta\in(0,1).\]
Notice, that $h_1(t)\leq h_2(t)$ implies $\Phi_{\eta,q}[h_1]\leq
\Phi_{\eta,q}[h_2].$
According to the ``power theorem'' of real interpolation theory,
see \cite{BL}, it holds
\begin{equation}\label{h1}
\Phi_{\eta,q}[K(\{u_n\},\cdot,p_0,p_1)]\asymp\|\{u_n\}\|_{\ell^{p,r}}^p,
\end{equation}
\begin{equation}\label{h2}
\Phi_{\eta,q}[K(f,\cdot,p_0,p_1)]\asymp\|f\|_{L^{p,r}}^p,
\end{equation}
\[
p=(1-\eta)p_0+\eta p_1,\ \ r=pq,\ \eta\in(0,1),\ 0<q\leq\infty,
\ 0<p_0,p_1<\infty,\ p_0\neq p_1.
\]
The quasi-norms on the right hand side denote the Lorentz scale of
sequence ideals $\ell^{p,r}$ or function spaces $L^{p,r},$
respectively. For the definition of these ideals see, e.g., \cite{BL}
or \cite{T}. We just point out, that $\ell_p=\ell^{p,p}$ and
$L_p=L^{p,p}.$

In general it is difficult to trace the constants in the two-side
estimates in \eqref{h1}, \eqref{h2}. However for the special case
$q=1$ one has the {\em equalities} (see \cite{BL}, p. 111,
proof of  Theorem 5.2.2.)
\begin{equation}\label{h3}
\Phi_{\eta,1}[K(\{u_n\},\cdot,p_0,p_1)]=\Theta(\eta,p_0,p_1)
\|\{u_n\}\|_{\ell_p}^p,
\end{equation}
\begin{equation}\label{h4}
\Phi_{\eta,1}[K(f,\cdot,p_0,p_1)]=\Theta(\eta,p_0,p_1)
\|f\|_{L_p}^p,
\end{equation}
\[
p=(1-\eta)p_0+\eta p_1,\ \eta\in(0,1),\ \ \ 0<p_0,p_1<\infty,\ p_0\neq p_1,
\]
where
\[\Theta(\eta,p_0,p_1)=\int_0^\infty t^{-\eta-1}\inf_{y_0+y_1=1}
(|y_0|^{p_0}+t|y_1|^{p_1})dt.
\]
Below we shall use these identities for improving the bounds on
$L_{\gamma,1}$ for certain $\gamma\in(1/2,3/2).$
\vspace{3mm}

{\bf 5.} In this subsection we consider the Schr\"odinger operator
\[H=-\Delta-V(x),\ \ V\geq 0,\ \ x\in{\Bbb R}^d,\]
in arbitrary dimensions $d\geq 1.$
We assume, that this operator is semibounded from below
and that its negative spectrum is discrete.
Let $\{E_n(H)\}$ be the non-decreasing sequence of negative eigenvalues
of the operator $H,$
each eigenvalue appears with its multiplicity.

Let us start from the Ky-Fan inequality for the discrete negative spectrum.
If $V=V_0+V_1$, and the operators
\[H_0=-\theta\Delta-V_0,\ \ H_1=-(1-\theta)\Delta-V_1,\ \ \theta \in(0,1),\]
have discrete negative spectrum , then the inequality
\[|E_{m+n-1}(H)|\leq|E_n(H_0)|+|E_m(H_1)|\]
holds for all $m,n=1,2...$ We construct the sequences
\[a_k:=E_s(H_0),\ \ \ s=1+\Bigl[\frac{k}{N+1}\Bigr] , \]
\[b_k:=E_l(H_1),\ \ \ l=N\Bigl[\frac{k}{N+1}\Bigr]+(k\mod N+1),\]
\[N,k,l,s\in {\Bbb N},\]
and obtain
\begin{equation}\label{hilfe1}
E_k(H)\leq a_k+b_k,\ \ \ \ k\in{\Bbb N}.
\end{equation}

Assume now $V_i\in L_{p_i+\varkappa}({\Bbb R}^d),\ \varkappa=d/2,
0<p_i<\infty$ for $d\geq 2$ and $1/2\leq p_i<\infty$ if $d=1.$
 From \eqref{hilfe1} and \eqref{LiTh} it follows, that
\[K(\{E_k(H)\},t,p_0,p_1)\leq \|\{a_k\}\|_{\ell_{p_0}}^{p_0}
+t\|\{b_k\}\|_{\ell_{p_1}}^{p_1}\leq\]
\[\leq (1+N)\sum_n |E_n(H_0)|^{p_0}+
t(1+N^{-1})\sum_m|E_m(H_1)|^{p_1}\]
\[\leq (1+N)\theta^{-\varkappa}L_{p_0,d}
\|V_0\|_{L_{p_0+\varkappa}({\Bbb R}^d)}^{p_0+\varkappa}
+t(1+N^{-1})(1-\theta)^{-\varkappa}L_{p_1,d}
\|V_1\|_{L_{p_1+\varkappa}({\Bbb R}^d)}^{p_1+\varkappa}.\]
Interchanging the definitions of the sequences $\{a_k\}$ and $\{b_k\}$
one can see, that in the previous expression
the role of $N$ and $1/N$ can be interchanged. Thus we can assume, that
$N$ is of the form $k$ or $1/k,\ k\in{\Bbb N}.$
Passing to the infimum over all
suitable decompositions $V=V_0+V_1$ one finds
\begin{equation}
K(\{E_k(H)\},t,p_0,p_1)\leq
\frac{(1+N)L_{p_0,d}}{\theta^\varkappa}
K(V,t\frac{(1+N^{-1})(1-\theta)^{-\varkappa}L_{p_1,d}}
{(1+N)\theta^{-\varkappa}L_{p_0,d}},p_0+\varkappa,p_1+\varkappa).
\end{equation}
\[N=..,\frac{1}{3},\frac{1}{2},1,2,3,..\ \ \varkappa=d/2,\]
with $0<p_i<\infty$ for $d\geq 2$ and $1/2\leq p_i<\infty$ for
$d=1.$
This relation allows one to apply interpolation methods {\em directly}
to the sequences of negative bound states, although the mapping
$V\mapsto\{E_n(H)\}$ is strongly non-linear.
\vspace{3mm}

{\bf 6.} Let us return to the one-dimensional case
and choose $p_0=1/2$ and $p_1=3/2.$
Applying the functional $\Phi_{\eta,1}$ to both sides of this
inequality, by \eqref{h3} and \eqref{h4} we obtain
\[\sum_k|E_k(H)|^\gamma
\leq L_{\gamma,1}^{**}\int V^{\gamma+\frac{1}{2}}dx,\ \ 1/2<\gamma<3/2,\]
where
\begin{equation}\label{**}
L_{\gamma,1}^{*,*}=
\frac{\Theta(\eta,1,2)}{\Theta(\eta,\frac{1}{2},\frac{3}{2})}
(1+N)^{1-\eta}\theta^{-(1-\eta)/2}L_{1/2,1}^{(1-\eta)}
(1+N^{-1})^\eta(1-\theta)^{-\eta/2}L_{p_1,1}^\eta,
\end{equation}
\[\gamma=(1-\eta)/2+3\eta/2,\ N=..,\frac{1}{3},\frac{1}{2},1,2,3,...\]
Let $M(\eta)$ be  the minimum of the sequence
\[(1+N)^{1-\eta}(1+N^{-1})^\eta,
\ N=...,\frac{1}{3},\frac{1}{2},1,2,3,...\]

It occurs, that
$M(\eta)\to 1$ as $\eta\to 0,1.$
If we minimize \eqref{**}
in $\theta\in(0,1),$ we find $\theta(\eta)=1-\eta$, and
\begin{equation}
L_{\gamma,1}^{*,*}= C(\eta)
\Bigl(\frac{\varsigma(3)}{3}\Bigr)^{(1-\eta)}
\Bigl(\frac{3}{16}\Bigr)^\eta,\ \ \ \gamma=\frac{1}{2}+\eta,
\end{equation}
\begin{equation}
C(\eta)=\frac{\Theta(\eta,1,2)}{\Theta(\eta,\frac{1}{2},\frac{3}{2})}
\frac{M(\eta)}
{\sqrt{\eta^\eta(1-\eta)^{1-\eta}}}.
\end{equation}
The involved functions $\Theta$ can be evaluated as
\[\Theta(\eta,1,2)=\frac{2^\eta}{\eta(1-\eta)(1+\eta)},\]
and
\[\Theta(\eta,\frac{1}{2},\frac{3}{2})=
\frac{\Bigl(\frac{2}{3}\sqrt{1+2/\sqrt{3}}\Bigr)^{1-\eta}}{1-\eta}
+\sqrt{\frac{1}{2}}\Bigl(\frac{3}{2}\Bigr)^\eta\Bigl(I_0(\eta)+
\frac{2}{3}I_1(\eta)\Bigr),\]
\[I_0(\eta)=\int_{u_0}^1 u(1-u)^{\frac{\eta-2}{2}}(1+u)^{\frac{\eta-1}{2}}du
\ ,\]
\[I_1(\eta)=\int_{u_0}^1 u(1-u)^{\frac{\eta}{2}}(1+u)^{\frac{\eta-3}{2}}du
\ ,\]
\[u_0=\sqrt{\frac{2}{2+\sqrt{3}}}\ \ .\]
Notice, that $C(\eta)\to 1$ as $\eta\to 1,$ thus
$L_{\gamma,1}^{**}\to 3/16$ as $\gamma\to 3/2.$
A numerical evaluation shows
$L_{\gamma,1}^{**}<L_{\gamma,1}^{*}$ as $\gamma>1.14 .$

\begin{tm} For the constant $L_{\gamma,1}$ in \eqref{upp1}
the bound $L_{\gamma,1}\leq\min\{L_{\gamma,1}^*,L_{\gamma,1}^{**}\},
\ 1/2<\gamma<3/2,$ holds.
\end{tm}
\vspace{3mm}

{\bf 7.} Let $\{\phi_i\}$ be some $L_2({\Bbb R}^d)-$orthonormal system,
$\phi_i\in W_2^1({\Bbb R}^d).$ Then  \eqref{LiTh} implies
(\cite{LT},\cite{Lieb})
\begin{equation}\label{ll1}
\sum_{i=1}^n\int|\nabla\phi_i|^2dx\geq K_{p,d}
\biggl(\int\rho_\phi^{p/(p-1)}dx\biggr)^{2(p-1)/d},
\end{equation}
\[\rho_\phi(x):=\sum_{i=1}^n|\phi_i(x)|^2,\]
\[\max\{d/2,1\}\leq p<1+d/2,\ \ \mbox{excluding}\ p=1\ \mbox{for}\ d=2,\]
with suitable constants $K_{p,d}.$
In case of $d=1$ and $p=3/2$ this turns into
\begin{equation}\label{ll}
\sum_{i=1}^n\int|\phi_i^\prime|^2dx\geq K_{3/2,1}\int\rho_\phi^{3}dx.
\end{equation}

The constant $K_{3/2,1}$ is related to $L_{1,1}$ via the formula
\[L_{1,1}=2/\sqrt{27 K_{3/2,1}}.\]
Our improved estimate on $L_{1,1}$ implies $K_1\geq 0.203,$
in compare with $K_1\geq 1/12$ in \cite{Lieb}.

We also point out the case $p=d=1.$
Then
\begin{equation}\label{ll2}
\sum_{i=1}^n\int|\phi_i^\prime|^2dx\geq
K_{1,1}\|\rho_\phi\|^2_{L_\infty({\Bbb R})} \ ,
\end{equation}
with a constant
$1\geq K_{1,1}\geq 1/(2L_{1/2,1}),$
see (3.27) in \cite{LT}.
Thus we find \eqref{ll2} with $1\geq K_{1,1}>0.497.$

\end{document}